\documentclass{JHEP3} % 10pt is ignored!

\usepackage{epsfig,multicol}
\usepackage{amsmath, amssymb}
\usepackage{concmath, palatino}

\newcommand\fverb{\setbox\pippobox=\hbox\bgroup\verb}
\newcommand\fverbdo{\egroup\medskip\noindent%
			\fbox{\unhbox\pippobox}\ }
\newcommand\fverbit{\egroup\item[\fbox{\unhbox\pippobox}]}
\newbox\pippobox
\newcommand{\be}{\begin{equation}} 
\newcommand{\ee}{\end{equation}}

\newcommand{\ba}{\begin{eqnarray}}
\newcommand{\ea}{\end{eqnarray}}

\newcommand{\la}{\longrightarrow}

\newtheorem{theorem}{Theorem}[section]

\newcommand{\ads}{AdS_5\times S^5}

\newcommand{\refeq}[1]{Eq.~(\ref{eq:#1})}

\title{Sum rules for higher twist $\mathfrak{sl}(2)$ operators \\
in ${\cal N}=4$ SYM}

\author{Matteo Beccaria and Francesca Catino\\
  Dipartimento di Fisica, Universita' del Salento,
  Via Arnesano, 73100 Lecce\\
  INFN, Sezione di Lecce\\
  E-mail: \email{matteo.beccaria@le.infn.it}\\
  E-mail: \email{francesca.catino@le.infn.it}}

\abstract{
The spectrum of anomalous dimensions of twist $\mathfrak{sl}(2)$ operators in ${\cal N}=4$ SYM
has an intriguing feature in low twist 2 or 3. The anomalous dimension of the lowest state, dual a folded string on $\ads$, 
can be computed by Bethe Ansatz at 3, 4 loops respectively as a simple closed function of the Lorentz spin. This feature is apparently lost
at higher twist. We propose sum rules for the excited anomalous dimensions where closed expressions 
can still be provided, even at higher twist. We present several explicit three loop examples. 
Many structural regularities can be observed leading to closed expressions which depend parametrically both on the spin and the 
twist. They allow to compute the subleading term in the logarithmic large spin expansion of the sum rules
as a compact simple function of the twist, in analogy with the recent results by Freyhult, Rej and Staudacher in arXiv:0712.2743 [hep-th].
}

\begin{document} 

\section{Introduction}
\label{sec:Intro}

The long range Bethe Ansatz solution of the mixing problem in ${\cal N}=4$ SYM 
allows to compute multi-loop anomalous dimensions of various composite operators in a very efficient 
way~\cite{Beisert:2005fw}. 
The main and unique obstacle is the celebrated {\em wrapping problem}~\cite{Ambjorn:2005wa} setting an upper bound
on the achievable order in the loop expansion. Within this bound, the exact perturbative
anomalous dimensions are recovered without approximation.

In some cases, we areinterested in parametric classes of operators where we would like to 
compute the spectrum as a closed function of the characterizing parameters (Lorentz spin, operator length, {\em etc.}).
This is a much more difficult question than merely asking the value of the anomalous dimensions at a specific point in the 
parameter space.

A positive answer beyond one-loop is not known in the general case, but is available for certain specific classes
of operators where additional insight saves the day. A very simple example is that of the generalized gluon condensate
\be
{\cal O}_L = \mbox{Tr}\,{\cal F}^L + \mbox{higher order mixing},
\ee
where ${\cal F}$ is a component of the self-dual field strength~\cite{Ferretti:2004ba,Beisert:2004fv,Rej:2007vm,Beccaria:2007gu}.
In this case, it is possible to derive compact five loop expressions for the anomalous dimension as a closed
function of the parameter $L$~\cite{Beccaria:2007gu}.

A much wider and more interesting class is that of quasi-conformal operators~\cite{Belitsky:2004cz}. 
They span an integrable sector of QCD and its various supersymmetric generalizations, including of course ${\cal N}=4$ SYM. 
In the cases relevant to our discussion, they take the form 
\be
{\cal O}_N = \sum_{n_1+\cdots+n_L} c_{n_1, \dots, n_L}\,\mbox{Tr} \big( D^{n_1} X\cdots D^{n_L} X \big),
\ee
where $D$ is a light-cone projected covariant derivative and 
$X$ can be an elementary scalar ($\varphi$) or a so-called good component of the gaugino ($\lambda$) or gauge ($A$) fields. In this 
context, $L$ is the twist of the operator and $N$ is the total Lorentz spin.

The case when $X$ is a scalar identifies the $\mathfrak{sl}(2)$ sector which is closed at all orders in the gauge coupling~\cite{Beisert:2003jj}. 
The gaugino case describes special operators in the purely fermionic closed $\mathfrak{sl}(2|1)$ subsector which evolve autonomously under
dilatations. Finally, the gauge operators close at one-loop and have a complicated pattern of higher order mixing.

In general, the one-loop classification of states is made simpler by the underlying collinear $\mathfrak{sl}(2)$ algebra under which the charges of 
$\varphi, \lambda, A$ are $s=\frac{1}{2}, 1, \frac{3}{2}$ respectively~\cite{Braun:2003rp}. 
The twist $L$ states belong to $[s]^{\otimes L}$ and can be 
decomposed in irreducible infinite dimensional $\mathfrak{sl}(2)$ modules. The twist-2 case is very special since supersymmetry links together
the three physical values of $s$~\cite{Belitsky:2003sh,Belitsky:2005gr}.
One has a single supermultiplets and a universal anomalous dimension $\gamma_{\rm univ}(N)$
describing (with trivial shifts in the Lorentz spin) all the states. This anomalous dimension can be computed at 3 loops
as a closed function of $N$ by invoking the  Kotikov, Lipatov, Onishchenko and Velizhanin (KLOV) 
principle leading to a simple Ansatz in terms of nested harmonic sums~\cite{klov}.

The case of twist-3 is more complicated. There are in principle three distinct series of modules associated to each elementary 
field~\cite{Beisert:2004di}.
Detailed results, including a nice application of superconformal symmetry, are described in~\cite{Beccaria:2007cn,Kotikov:2007cy,Beccaria:2007bb} for
the scalar sector, \cite{Beccaria:2007vh} for the gaugino sector, and \cite{Beccaria:2007pb,Beccaria:2008fi} for the gluon sector.

A curious fact is that the ground state (i.e. that with smallest anomalous dimension) admits {\em again} closed expressions in all sectors
based on suitably generalized KLOV-like principles.
As soon as one moves to higher twist, even in the simplest case of $X = \varphi$, it is easy to check that no simple closed expressions 
describe the ground state whose anomalous dimension is irrational. Therefore, the following basic questions naturally arise:
\begin{enumerate}
\item Why are twist-2 and 3 so special ? 
\item Can we generalize the twist-2 and 3 closed expressions to higher twists ?
\end{enumerate}

The aim of this paper is precisely that of answering the above questions. We shall show that, at higher twist, it is possible to 
consider sums of (powers of) anomalous dimensions of the ground and excited states. For these combinations, we shall provide quite simple
closed formulas as well as compact sum rules which are parametric in both $N$ and the twist $L$

These expressions are a simple hidden constraint on the anomalous dimension whose precise nature is not clear. In particular, they suggest 
similar sum rules at strong coupling for the dual string states discussed for instance in~\cite{Frolov:2002av,Kruczenski:2006pk,Ishizeki:2007we,Ishizeki:2007kh,Kruczenski:2008bs}.

The plan of the paper is the following. 
In Sec.~(\ref{sec:xxx}) we recall some basic facts concerning the $XXX_{-s}$ integrable spin chain and provide several explicit 
examples in Sec.~(\ref{sec:xxx-examples}). The outcome of this analysis is summarized in Sec.~(\ref{sec:lessons}). In Sec.~(\ref{sec:log})
we recall a few important properties of the large spin expansion of twist anomalous dimensions. 
In Sec.~(\ref{sec:linear}) we propose linear sum rules for the singlet anomalous dimensions of various twist operators.
In Sec.~(\ref{sec:linear-scalar}) we present three loop results for these sum rules in the scalar sector. These results suggest a twist-dependent
conjecture formulated in Sec.~(\ref{sec:linear-scalar-twist}). Various checks are performed in Sec.~(\ref{sec:linear-checks}). The subleading 
corrections at large spin are computed in Sec.~(\ref{sec:linear-subleading}). Quadratic sum rules are proposed in Sec.~(\ref{sec:quadratic}), 
elaborated in Sec.~(\ref{sec:quadratic-twist}) and checked in Sec.~(\ref{sec:quadratic-checks}). A few results for cubic sum rules are finally
presented in Sec.~(\ref{sec:cubic}). Various Appendices are devoted to some technical results.

\section{The integrable $XXX_{-s}$ chain}
\label{sec:xxx}

In this section, we briefly recall a few basic facts about the integrable $XXX_{-s}$ spin chain. It
 describes the one-loop mixing of twist-$L$ quasipartonic operators built with elementary collinear conformal spin $s$ fields.
A nice and accessible recent review on this standard material is~\cite{Derkachov:1999pz}. Our presentation is brief and just sets up the language for 
the later discussion.

\subsection{Basic facts}

The $XXX_{-s}$ chain is a quantum spin chain with $\mathfrak{sl}(2, \mathbb{R})$ symmetry. 
Each site carries the infinite dimensional $[s]$ representation of the $SL(2, \mathbb{R})$ collinear conformal group.
The decomposition rule for the tensor products of this representation reads
\be
\label{eq:decomp}
[s]\otimes [s] = \bigoplus_{n=0}^\infty [s+n],
\ee
and can be used to analyze the states of a chain with $L$ sites.
The local  spin chain integrable Hamiltonian reads
\be
H = \sum_{n=1}^L H_{n, n+1},
\ee
with 
\be
H_{n, n+1} = \psi(J_{n, n+1})-\psi(2\,s),\qquad \psi(z) = \frac{d}{dz}\,\log\Gamma(z).
\ee
The quantity $J_{n, n+1}$ is the spin of the two-site states 
\be
(\vec S_n + \vec S_{n+1})^2 = J_{n, n+1}\,(J_{n, n+1}-1).
\ee
This Hamiltonian is integrable and can be studied by Bethe Ansatz or Baxter techniques.

\subsection{Relation with ${\cal N}=4$ SYM operators and anomalous dimensions}

The $XXX_{-s}$ Hamiltonian is the mixing matrix for the one-loop renormalization of planar composite operators of the form 
\be
{\cal O}_L(N) = \sum_{n_1+\cdots+n_L=N} c_{n_1,\dots, n_L}\,\mbox{Tr}\left\{
D_+^{n_1}X\cdots D_+^{n_L}X\right\},
\ee
where $s=1/2, 1, 3/2$ for the physical cases $X = \varphi, \lambda, A$. A straightforward application of \refeq{decomp}
leads to 
\be
[s]^{\otimes L} = \bigoplus_{N=0}^\infty g_L(N) \left[\frac{L}{2}+N\right], \qquad
g_L(N) = \binom{N+L-1}{L-1},
\ee
which tells that there are $g_L(N)$ highest weight states with Lorentz spin $N$. These are in 1-1 correspondence with the non-trivial ({\em i.e.} without
roots at infinity) solutions of the Bethe Ansatz equations (BAE)
\be
\left(\frac{u_k+i\,s}{u_k-i\,s}\right)^L = \mathop{\prod_{j=1}^N}_{j\neq k}\frac{u_k-u_j-i}{u_k-u_j+i}.
\ee
A short calculation also provides the energy and momentum charges
\be
E = \sum_{k=1}^N\frac{2\,s}{u_k^2+s^2},\qquad e^{i\,P} = \prod_{k=1}^N\frac{u_k-i\,s}{u_k+i\,s}.
\ee
These energies are the one-loop anomalous dimensions of the above operators, say
\be
\gamma = \frac{\lambda}{8\,\pi}\,E,
\ee
where $\lambda$ is the 't Hooft large $N_c$ coupling.

Clearly, it is complicated to enumerate the full set of solutions to the BAE. 
A useful alternative method is the Baxter approach described in the next section.

\subsection{The Baxter equation for the $XXX_{-s}$ chain}

An alternative approach to the solution of the BAE is based on the Baxter approach~\cite{Bax72}.
The main tool is the Baxter operator whose eigenvalues $Q(u)$ obey a relatively simple functional equation.
If $Q(u)$ is assumed to be a polynomial, then the Baxter equation is equivalent to the algebraic Bethe Ansatz equations for its roots
to be identified with the Bethe roots. A more general discussion can be found 
in~\cite{Derkachov:1999pz,KorTrick1}.

In practice, the Baxter approach in the present case is quite simply stated. One introduces the Baxter function which is the minimal 
polynomial with roots equal to the Bethe roots
\be
Q(u) = \prod_{k=1}^N (u-u_k).
\ee
The BA equations are equivalent to the Baxter equation that we write for general conformal spin $s$
(although we shall be mainly interested in the case $s=1/2$)
\be
(u+i\,s)^L\,Q(u+i)+(u-i\,s)^L\,Q(u-i) = t(u)\,Q(u).
\ee
where 
\ba
t(u) &=& 2\,u^L + q_2\,u^{L-2} + q_3\,u^{L-3} + \cdots + q_L, \\
q_2 &=& -(N+L\,s)\,(N+L\,s-1)+L\,s\,(s-1).
\ea
The quantities $q_3, \dots, q_L$ have the meaning of quantum numbers. They must be obtained by consistence of the 
Baxter equation and the assumption of a polynomial Baxter function.
Once $Q$ is found, the energy and momentum can be written in terms of $Q$ as 
\be
E = i\,\left[(\log\,Q(u))'\right]^{+i\,s}_{-i\,s},\qquad e^{i\,P} = \frac{Q(+i\,s)}{Q(-i\,s)}.
\ee

\section{The $XXX_{-s}$ chain at twist $L=2, \dots, 5$}
\label{sec:xxx-examples}

In order to set the stage for the later discussion and introduction of sum rules, we now illustrate very explicitly
the structure of the Baxter equation at various small twists $L=2, \dots, 5$. Our results will be generically valid 
positive values $s>0$ of the (quantized) conformal spin.

\subsection{Twist 2}

For $L=2$ and Lorentz spin $N$, the Baxter equation is 
\be
(u+i\,s)^2\,Q(u+i)+(u-i\,s)^2\,Q(u-i) = t(u)\,Q(u),
\ee
with 
\ba
t(u) &=& 2\,u^2 + q_2, \\
q_2 &=& N-N^2-4\,N\,s-2\,s^2. 
\ea
In this case, there are no additional quantum numbers. This follows from the trivial multiplicities in 
\be
[s]\otimes [s] = \bigoplus_{N=0}^\infty  [2\,s+N].
\ee

\medskip
The Baxter polynomial with degree $N$ (even or odd) is~\cite{Derkachov:2002tf}
\be
Q(u) = {}_3 F_2\left(\left. \begin{array}{c}
-N\quad N+4\,s-1\quad s-i\,u \\
2\,s\quad 2\,s
\end{array}
\right| 1\right).
\ee
The Baxter polynomial has parity
\be
Q(-u) = (-1)^N\,Q(u)\qquad\la\qquad e^{i\,P} = (-1)^N.
\ee
The energy is 
\be
E = i\,\left[(\log\,Q(u))'\right]^{+i\,s}_{-i\,s} = 4\,\left[\psi(N+2\,s)-\psi(2\,s)\right],
\ee
where
\be
\psi(z) = \frac{d}{dz}\,\log\,\Gamma(z).
\ee

\subsection{Twist 3}

For $L=3$ and spin $N$, the Baxter equation is 
\be
(u+i\,s)^3\,Q(u+i)+(u-i\,s)^3\,Q(u-i) = t(u)\,Q(u),
\ee
with 
\ba
t(u) &=& 2\,u^3 + q_2\,u + q_3, \\
q_2 &=& N-N^2-6\,N\,s-6\,s^2. 
\ea
In this case, there is an additional quantum number. This is related to the multiplicities in 
\be
[s]\otimes [s]\otimes [s] = \bigoplus_{n_1, n_2=0}^\infty  [3\,s+n_1+n_2] = \bigoplus_{N=0}^\infty  (N+1)\,[3\,s+N].
\ee
Indeed, looking for a polynomial solution to the Baxter equation we find the condition
\be
P(q_3)=0,\qquad \mbox{deg}\,P = N+1.
\ee
Let us analyze the solutions for real $s>0$.

\medskip
\noindent
\underline{\bf even $N$}
\medskip

\noindent
For even $N$, the polynomial $P$ reads
\be
P(q_3) = q_3\, R(q_3),\qquad R(q_3) = R(-q_3).
\ee
Hence we can have $q_3=0$ or $q_3 = \pm q^*$ for some values of $q^*\neq 0$. The Baxter function associated with $q_3=0$ is even 
and is associated with a zero momentum state which turns out to be the ground state.

The explicit form of $Q(u)$ is known and reads
\be
Q(u) = {}_4 F_3\left(\left. \begin{array}{c}
-\frac{N}{2}\quad \frac{N}{2}+3\,s-\frac{1}{2}\quad \frac{1}{2}+i\,u\quad \frac{1}{2}-i\,u \\
\frac{1}{2}+s\quad\frac{1}{2}+s\quad\frac{1}{2}+s
\end{array}
\right| 1\right) = Q(-u).
\ee
To compute the energy, it is convenient to relate $Q$ to the Wilson polynomials
\be
\frac{W_n(u^2, a, b, c, d)}{(a+b)_n\,(a+c)_n\,(a+d)_n} = {}_4 F_3\left(\left.
\begin{array}{c}
-n\quad n+a+b+c+d-1\quad a+i\,u\quad a-i\,u\\
a+b\quad a+c\quad a+d
\end{array}
\right| 1\right).
\ee
We find apart from a trivial scaling
\be
Q(u) = W_{N/2}\left(u^2, s, s, s, \frac{1}{2}\right).
\ee
Using the following formula from the Appendix B of \cite{Derkachov:1999ze}
\ba
\lefteqn{\left. i\,\frac{d}{du}\,W_n(u^2, a, a, c, d) \right|_{u = i\,a} =} &&  \\
&& = \psi(n+a+c)-\psi(a+c)+\psi(n+a+d)-\psi(a+d)
\ea
and the fact that $W$ is invariant under permutations of $a, b, c, d$, we find the result
\ba
\lefteqn{E = i\,\left[(\log\,Q(u))'\right]^{+i\,s}_{-i\,s} =} && \\
&& 2\,\left[\psi\left(\frac{N}{2}+2\,s\right)-\psi(2\,s)+\psi\left(\frac{N}{2}+s+\frac{1}{2}\right)-\psi\left(s+\frac{1}{2}\right)\right].
\ea
Notice that for the interesting values $s=1/2, 1, 3/2$ we can simplify the resulting expressions and find
\ba
s = \frac{1}{2},\qquad && E(N) = 4\,\left[\psi\left(\frac{N}{2}+1\right)-\psi(1)\right], \\
s = 1, \qquad && E(N) = 4\,\left[\psi(N+3)-\psi(3)\right], \\
s = \frac{3}{2}, \qquad && E(N) = 2\,\left[\psi\left(\frac{N}{2}+3\right)+\psi\left(\frac{N}{2}+2\right)-\psi(3)-\psi(2)\right].
\ea
The paired states with $q_3 = \pm q^*$ are associated with degenerate values of the energy. Some of them can have zero momentum.

An example is the case $N=6$ at $s=\frac{1}{2}$. The Baxter function associated with $q_3=0$ is 
\be
Q(u) = u^6-\frac{19 u^4}{4}+\frac{323 u^2}{80}-\frac{153}{320},\qquad \gamma = \frac{22}{3},\qquad e^{i\,P}=1.
\ee
There are other two solutions with $P=0$ which are obtained with $q_3 = \pm 2\,\sqrt{723}$. The associated Baxter functions
are related by parity. One of them reads
\be
Q(u) = u^6+\frac{2 \sqrt{723} u^5}{13}+\frac{235 u^4}{52}+\frac{5}{143} \sqrt{\frac{241}{3}} u^3-\frac{2523 u^2}{2288}-\frac{23 \sqrt{\frac{241}{3}}
   u}{1144}+\frac{155}{9152},
\ee
and has 
\be
\gamma = \frac{227}{20},\qquad e^{i\,P}=1.
\ee

\medskip
\noindent
\underline{\bf odd $N$}
\medskip

\noindent
For even $N$, the polynomial $P$ is even
\be
P(q_3) = P(-q_3).
\ee
Hence we have only $q_3 = \pm q^*$ associated with degenerate states with Baxter functions related by $u\to -u$. 
Again, some of these states can have zero momentum.

An example is the case $N=3$ at $s=\frac{1}{2}$. There are two paired zero momentum states with $q_3 = \pm\frac{3}{2}\sqrt{35}$ and 
Baxter function (the other is related by parity)
\be
Q(u) = u^3+\frac{3}{2} \sqrt{\frac{5}{7}} u^2+\frac{u}{4}-\frac{1}{8 \sqrt{35}},\qquad \gamma = \frac{15}{2},\qquad e^{i\,P}=1.
\ee

\subsection{Twist 4}

For $L=4$ and spin $N$ the Baxter equation is 
\be
(u+i\,s)^4\,Q(u+i)+(u-i\,s)^4\,Q(u-i) = t(u)\,Q(u),
\ee
with 
\ba
t(u) &=& 2\,u^4 + q_2\,u^2 + q_3\,u+q_4, \\
q_2 &=& N-N^2-8\,N\,s-12\,s^2. 
\ea
In this case, there are two quantum numbers. They must agree with the multiplicities in 
\be
[s]\otimes [s]\otimes [s]\otimes [s] = \bigoplus_{n_1, n_2, n_3=0}^\infty  [4\,s+n_1+n_2+n_3] = \bigoplus_{N=0}^\infty  \frac{(N+1)(N+2)}{2}\,[4\,s+N].
\ee
If $N$ is even, looking for a polynomial solution to the Baxter equation we find the conditions
\ba
P(q_3, q_4) &=& 0,\\
q_3\,R(q_3, q_4) &=& 0.
\ea
If $q_3=0$, we find
\be
P(0, q_4) \equiv S(q_4) = 0,\qquad \mbox{deg} S = \frac{N}{2}+1.
\ee
These are non-degenerate states with $Q(u)=Q(-u)$ hence zero momentum. Notice that in this case the transfer matrix
is even, a property which is related to the parity invariance of $Q$.

The solutions with $q_3\neq 0$ appear in degenerate pairs and can have zero momentum.
If $N$ is odd, looking for a polynomial solution to the Baxter equation we find again  conditions
\ba
P(q_3, q_4) &=& 0,\\
q_3\,R(q_3, q_4) &=& 0.
\ea
If $q_3=0$, we find solutions with $Q(u)=-Q(-u)$ hence non-zero momentum. 
The solutions with $q_3\neq 0$ appear in degenerate pairs and can have zero momentum.

\subsection{Twist 5}

For $L=5$ and spin $N$ the Baxter equation is 
\be
(u+i\,s)^5\,Q(u+i)+(u-i\,s)^5\,Q(u-i) = t(u)\,Q(u),
\ee
with 
\ba
t(u) &=& 2\,u^5 + q_2\,u^3 + q_3\,u^2+q_4\,u+q_5, \\
q_2 &=& N-N^2-10\,N\,s-20\,s^2. 
\ea
In this case, there are three quantum numbers. They must agree with the multiplicities in 
\be
[s]^{\otimes 5} = \bigoplus_{N=0}^\infty  \frac{(N+1)(N+2)(N+3)}{6}\,[5\,s+N].
\ee
We focus on the non-degenerate states. These are present for even $N$ and when the transfer matrix has definite parity. In this case, this means
\be
t(u) = 2\,u^5 + q_2\,u^3 + q_4\,u.
\ee
The Baxter equation reduces to a polynomial in $q_4$ that turns out to have degree
\be
P(q_4) = 0,\qquad \mbox{deg}\,P = \frac{N}{2}+1,
\ee
as in the $L=4$ case.

\section{Lessons from the previous analysis and general features}
\label{sec:lessons}

Let us stop to illustrate a few important features emerging from the previous long and explicit discussion.

\begin{enumerate}
\item The case $L=2$ is well known. There is a single $\mathfrak{sl}(2)$ highest state for each spin. It has the correct zero
momentum only for even spin.

\item The case $L=3$ is also rather well known~\cite{Beccaria:2007cn,Kotikov:2007cy,Beccaria:2007bb,Beccaria:2007vh,Beccaria:2007pb}.
We consider only the zero momentum states which are the only relevant ones to planar ${\cal N}=4$ SYM.
For even spin, the minimal energy state is a singlet. The other {\em excited states} appear always in degenerate pairs.
For odd spin, there are no singlet states. 
Degenerate states are associated with a symmetry in the planar limit relating traces with reversed traces~\cite{Beisert:2003tq}
\be
\mbox{Tr}(D^{n_1} \varphi\cdots D^{n_L}\varphi) \leftrightarrow \mbox{Tr}(D^{n_L} \varphi\cdots D^{n_1}\varphi).
\ee

\item For $L>3$, the (zero momentum) highest weight states can be divided into two subsets. {\em Singlets} with non degenerate energy, and {\em Paired} states with 
degeneracy 2. 
\end{enumerate}
From the symmetry of the Baxter equation it is easy to proof the
\begin{theorem} The singlets are all obtained by solving the Baxter equation with the requirement that the transfer matrix eigenvalue 
$t(u)$ has definite parity
\be
t(-u) = (-1)^L\,t(u). 
\ee
\end{theorem}
This sets to zero several quantum numbers. Let us relabel the remaining free quantum numbers (conserved charges) as $z_i$. The pattern is 
clear from the following list of definite parity transfer matrices (remember that $q_2$ is known)
\ba
t_3(u) &=& 2\,u^3 + q_2\,u, \\
t_4(u) &=& 2\,u^4 + q_2\,u^2 + z_1, \\
t_5(u) &=& 2\,u^5 + q_2\,u^3 + z_1\,u, \\
t_6(u) &=& 2\,u^6 + q_2\,u^4 + z_1\,u^2 + z_2, \\
t_7(u) &=& 2\,u^7 + q_2\,u^5 + z_1\,u^3 + z_2\,u.
\ea
The number of singlet states for a certain twist is the number of possible values of these quantum numbers. It is a function of the Lorentz spin
given by the following simple formula
\be
L = 2\,n, 2\,n+1,\qquad \#\, \mbox{singlets} = \binom{\frac{N}{2}+n-1}{n-1}.
\ee
For illustration, we show in Figs.~(\ref{fig:spectrum3}) and (\ref{fig:spectrum4}) the full spectrum of highest weight states 
at $L=3, 4$. A  general feature is that the singlet part of the spectrum embraces the full spectrum. 
In particular, the lowest and highest states are singlets.

\section{Logarithmic scaling of the anomalous dimensions}
\label{sec:log}

The following general information is known about the band of highest weight anomalous dimensions at generic twist $L$.
We focus on the scalar $s=1/2$ sector, but generalizations are possible.
In the $N\to\infty$ limit, the minimal anomalous dimension has a logarithmic scaling which is twist independent and reads
\be
\label{eq:logscaling}
\gamma_{\rm min}\sim f(g)\,\log\,N.
\ee
The most explicit proofs of this statement in ${\cal N}=4$ SYM are~\cite{Korchemsky:1992xv,Belitsky:2003ys,Belitsky:2006en} 
and~\cite{Eden:2006rx,Beisert:2006ez}.
The {\em scaling function} $f(g)$ is proportional to the physical coupling, {\em a.k.a.} cusp 
anomalous dimension~\cite{Korchemsky:1992xv} and has been computed at all orders in~\cite{Beisert:2006ez}. At three loops, it reads
\ba
f(g) &=& 4\,g^2_{\rm ph}, \\
g^2_{\rm ph} &=& g^2 -\zeta_2\,g^4 + \frac{11}{5}\,\zeta_2^2\,g^6 + \cdots = \nonumber \\
&=& g^2 -\frac{\pi^2}{6}\,g^4 + \frac{11\,\pi^4}{180}\,\,g^6 + \cdots~.
\ea
In twist-2, the physical principle behind the scaling \refeq{logscaling} is simply that the large $N$ limit is nothing but the 
quasi-elastic $x_{\rm Bjorken}\to 1$ deep inelastic scattering regime. This is dominated by universal classical soft gluon emission
characterized by the anomalous cusp contribution of the quasi-free parton dynamics.

The {\em excited} anomalous dimensions are expected to scale in the same way but with a different prefactor ranging as follows
\be
f(g)\,\log\,N < \gamma < \frac{L}{2}\,f(g)\,\log\,N .
\ee
We remark that this can be nicely understood, at strong coupling, in terms of the dual string configurations which have $L$ spikes each contributing $\frac{1}{2}\,f(g)$ to the 
coefficient in the case where they are equally spaced~\cite{Kruczenski:2008bs}.

At next-to-leading logarithmic order, a general formula has been recently derived in~\cite{Freyhult:2007pz} for the minimal 
anomalous dimension at generic twist. It reads
\be
\label{eq:generalized1}
\gamma_{\rm min} =  f(g)\,\log\,N + f_{\rm sl}(g, L) + \mbox{suppressed terms},
\ee
where, in our notation for the coupling
\ba
\label{eq:generalized2}
f_{\rm sl}(g, L) &=& (\gamma_E-(L-2)\,\log\,2)\,f(g)-2\,(7-2\,L)\,\zeta_3\,g^4 + \\
&& + \left(-\frac{L-4}{3}\,\pi^2\,\zeta_3 + (62-21\,L)\,\zeta_5\right)\,g^6 + \cdots~. \nonumber
\ea
This formula is remarkable because it gives the explicit twist dependence, reabsorb the scaling function  in a compact way and
provides the other corrections as $\zeta$-terms with simple twist dependence.

Given the above general constraints, is it possible to explore analytically the spectrum of highest weights at general 
twist ? A partially positive answer is provided by sum rules that we know describe.

\section{Linear sum rules at one-loop}
\label{sec:linear}

Let $\gamma^{(s)}_{L,k}(N)$ denote the anomalous dimensions of the various highest states, labeled by $k$.
We consider at one-loop all the three physical values $s=1/2, 1, 3/2$ associated with elementary scalars, gauginos, and gauge fields. 
We can compute the sum of the anomalous dimensions of singlet states
\be
\Sigma_L^{(s)}(N) = \sum_{k\in \ \rm singlets}\gamma_{L, k}^{(s)}(N).
\ee
It turns out that this quantity is {\bf rational} ! The reason is very simple. The above sum can be computed by the Baxter approach.
The anomalous dimensions are given by a rational function of the free charges which are not fixed by the parity constraint on the 
transfer matrix. These charges are constrained and fully determined by a system of polynomial equations. So, the point is to show that 
the sum of a rational function over the roots of a system of (rational) polynomials is rational. This simple theorem is proved and 
discussed in App.~(\ref{app:rationality}).

Given a sequence of rational numbers describing the $N$ dependence of $\Sigma_L^{(s)}(N)$, it is possible to look for closed
formulas, by using some trial and error combination of harmonic sums, as inspired by the low $L$ cases. Our notation for 
harmonic sums is standard ($a\in \mathbb{Z}$, $\mathbf{a} = (a_1, \dots, a_n)$ with $a_i\in\mathbb{Z}$)
\be
S_a(N) = \sum_{k=1}^N\frac{(\mbox{sign}\,a)^k}{k^{|a|}}, \qquad S_{a, \mathbf{b}}(N) = 
\sum_{k=1}^N\frac{(\mbox{sign}\,a)^k}{k^{|a|}}\,S_\mathbf{b}(k).
\ee
The following closed formulae are obtained for the cases $L=4, 5, 6, 7$ (they are fulfilled by any $N$ we have been able to test, typically of the
order ${\cal O}(100)$)
\ba
L=4, 5 &\qquad& \Sigma_L^{(s)}(N) = \sum_{n=1}^{\frac{N}{2}}\left[\sigma_L^{(s)}(n)-\sigma_L^{(s)}(0)\right],\\
L=6, 7 &\qquad& \Sigma_L^{(s)}(N) = \sum_{n=1}^{\frac{N}{2}}\,\sum_{m=1}^n \left[\sigma_L^{(s)}(m)-\sigma_L^{(m)}(0)\right],
\ea
where for $L=4$
\ba
\sigma_4^{(1/2)}(n) &=& 6\,S_1(n)+2\,S_1(2\,n-1)-2\,S_{-1}(2\,n-1), \\
\sigma_4^{(1)}(n)   &=& 2\,S_1(n)+4\,S_1(n+1)+2\,S_1(2\,n+1)-2\,S_{-1}(2\,n+1), \\
\sigma_4^{(3/2)}(n) &=& 2\,S_1(n+1)+4\,S_1(n+2)+2\,S_1(2\,n+1)-2\,S_{-1}(2\,n+1),
\ea
for $L=5$
\ba
\sigma_5^{(1/2)}(n) &=& 8\,S_1(n),\\
\sigma_5^{(1)}(n)   &=& 6\,S_1(n+1)+2\,S_1(2\,n+1)-2\,S_{-1}(2\,n+1), \\
\sigma_5^{(3/2)}(n) &=& 2\,S_1(n+1)+6\,S_1(n+2),
\ea
for $L=6$
\ba
\sigma_6^{(1/2)}(n) &=& 10\,S_1(n)+2\,S_1(2\,n-1)-2\,S_{-1}(2\,n-1), \\
\sigma_6^{(1)}(n)   &=& 2\,S_1(n)+8\,S_1(n+1)+2\,S_1(2\,n+1)-2\,S_{-1}(2\,n+1), \\
\sigma_6^{(3/2)}(n) &=& 2\,S_1(n+1)+8\,S_1(n+2)+2\,S_1(2\,n+1)-2\,S_{-1}(2\,n+1),
\ea
and for $L=7$
\ba
\sigma_7^{(1/2)}(n) &=& 12\,S_1(n),\\
\sigma_7^{(1)}(n)   &=& 10\,S_1(n+1)+2\,S_1(2\,n+1)-2\,S_{-1}(2\,n+1), \\
\sigma_7^{(3/2)}(n) &=& 2\,S_1(n+1)+10\,S_1(n+2).
\ea
Notice that for $s=1/2$ we can also write in a more uniform way
\ba
\sigma_4^{(1/2)}(n) &=& 8\,S_1(2\,n)+4\,S_{-1}(2\,n), \\
\sigma_5^{(1/2)}(n) &=& 8\,S_1(n), \\
\sigma_6^{(1/2)}(n) &=& 12\,S_1(2\,n)+8\,S_{-1}(2\,n), \\
\sigma_7^{(1/2)}(n) &=& 12\,S_1(n).
\ea
where we have exploited the remarkable identity
\be
S_1(2\,s-1)-S_{-1}(2\,s-1) = 2\,S_{-1}(2\,s)+4\,S_1(2\,s)-3\,S_1(s),\qquad s\in 2\,\mathbb{N}.
\ee

\section{Linear sum rules at three loop results in the scalar sector}
\label{sec:linear-scalar}

Starting from the one-loop Bethe roots evaluated with the Baxter approach, one can build the multi-loop anomalous dimensions
by feeding the long-range Bethe equations of~\cite{Beisert:2005fw}. This is quite easy in the $s=1/2$ case
where the Bethe equations are not nested. The procedure is standard (see for instance the detailed discussion in~\cite{Kotikov:2007cy}). From a long list for several even values 
of $N$, one 
makes an Ansatz with higher transcendentality nested sums and solves the over constrained system of equations.
Dropping for simplicity the $s=1/2$ label and denoting
\be
\sigma_L(n) = \sum_{\ell\ge 1} g^{2\,\ell} \sigma_{L, \ell}(n),
\ee
one finds the following solutions.

\medskip
\noindent
\underline{\bf L=4}

\medskip
\noindent
The argument of the harmonic sums is 
\be
S_{\cdots} \equiv S_{\cdots}(2\,n).
\ee
\ba
\sigma_{4, 2} &=& 16 \,S_{-3}+24 \,S_3-16 \,S_{-2,-1}-8 \,S_{-2,1}-8 \,S_{-1,-2}-8 \,S_{-1,2}-16 \,S_{1,-2} \nonumber\\
&& -16 \,S_{1,2}-16 \,S_{2,-1}-24 \,S_{2,1}, \\
\sigma_{4, 3}^{(1/2)} &=& 104 \,S_{-5}+184 \,S_5-144 \,S_{-4,-1}-104 \,S_{-4,1}-192 \,S_{-3,-2}-112 \,S_{-3,2}-144 \,S_{-2,-3} \nonumber\\
&& -96 \,S_{-2,3}-96 \,S_{-1,-4}-96 \,S_{-1,4}-192 \,S_{1,-4}-192 \,S_{1,4}-184
   \,S_{2,-3}-248 \,S_{2,3}-176 \,S_{3,-2} \nonumber\\
&& -288 \,S_{3,2}-144 \,S_{4,-1}-216 \,S_{4,1}+64 \,S_{-3,-1,-1}+128 \,S_{-3,-1,1}+64 \,S_{-3,1,-1}+32 \,S_{-3,1,1} \nonumber\\
&& +64 \,S_{-2,-2,-1}+96
   \,S_{-2,-2,1}+64 \,S_{-2,-1,-2}+64 \,S_{-2,-1,2}+32 \,S_{-2,1,-2}+32 \,S_{-2,1,2} \nonumber\\
&& +64 \,S_{-2,2,-1}+32 \,S_{-2,2,1} +64 \,S_{-1,-3,-1}+64 \,S_{-1,-3,1}+32 \,S_{-1,-2,-2}+32
   \,S_{-1,-2,2}\nonumber\\
&& +32 \,S_{-1,2,-2}+32 \,S_{-1,2,2}+64 \,S_{-1,3,-1}+64 \,S_{-1,3,1}+128 \,S_{1,-3,-1}+128 \,S_{1,-3,1}+64 \,S_{1,-2,-2}\nonumber\\
&& +64 \,S_{1,-2,2}+64 \,S_{1,2,-2}+64 \,S_{1,2,2}+128
   \,S_{1,3,-1}+128 \,S_{1,3,1}+128 \,S_{2,-2,-1}+80 \,S_{2,-2,1}\nonumber\\
&& +64 \,S_{2,-1,-2}+64 \,S_{2,-1,2}+96 \,S_{2,1,-2}+96 \,S_{2,1,2}+128 \,S_{2,2,-1}+160 \,S_{2,2,1}+128 \,S_{3,-1,-1}\nonumber\\
&& +64
   \,S_{3,-1,1}+128 \,S_{3,1,-1}+192 \,S_{3,1,1}~.
\ea

\medskip
\noindent
\underline{\bf L=5}

\medskip
\noindent
The argument of the harmonic sums is 
\be
S_{\cdots} \equiv S_{\cdots}(n).
\ee
\ba
\sigma_{5, 2} &=& 8 \,S_3-8 \,S_{1,2}-12 \,S_{2,1}, \\
\sigma_{5, 4} &=& 21 \,S_5-24 \,S_{1,4}-38 \,S_{2,3}-46 \,S_{3,2}-36 \,S_{4,1}+16 \,S_{1,2,2}+32 \,S_{1,3,1}\nonumber\\
&& +24 \,S_{2,1,2}+40 \,S_{2,2,1}+48 \,S_{3,1,1}~.
\ea

\medskip
\noindent
\underline{\bf L=6}

\medskip
\noindent
The argument of the harmonic sums is 
\be
S_{\cdots} \equiv S_{\cdots}(2\,n).
\ee
\ba
\sigma_{6, 2} &=& 40 \,S_{-3}+48 \,S_3-32 \,S_{-2,-1}-24 \,S_{-2,1}-16 \,S_{-1,-2}-16 \,S_{-1,2}-24 \,S_{1,-2}\nonumber\\
&& -24 \,S_{1,2}-32 \,S_{2,-1}-40 \,S_{2,1}, \\
\sigma_{6, 3} &=&  576 \,S_{-5}+736 \,S_5-480 \,S_{-4,-1}-384 \,S_{-4,1}-560 \,S_{-3,-2}-448 \,S_{-3,2}-400 \,S_{-2,-3} \nonumber\\
&& -352 \,S_{-2,3}-192 \,S_{-1,-4}-192 \,S_{-1,4}-288 \,S_{1,-4}-288 \,S_{1,4}-464
   \,S_{2,-3}-512 \,S_{2,3} \nonumber\\
&& -560 \,S_{3,-2}-672 \,S_{3,2}-480 \,S_{4,-1}-576 \,S_{4,1}+256 \,S_{-3,-1,-1}+320 \,S_{-3,-1,1}\nonumber\\
&& +256 \,S_{-3,1,-1}+192 \,S_{-3,1,1}+192 \,S_{-2,-2,-1}+224
   \,S_{-2,-2,1}+128 \,S_{-2,-1,-2}+128 \,S_{-2,-1,2}\nonumber\\
&& +96 \,S_{-2,1,-2}+96 \,S_{-2,1,2}+192 \,S_{-2,2,-1}+160 \,S_{-2,2,1}+128 \,S_{-1,-3,-1}+128 \,S_{-1,-3,1}\nonumber\\
&& +64 \,S_{-1,-2,-2}+64
   \,S_{-1,-2,2}+64 \,S_{-1,2,-2}+64 \,S_{-1,2,2}+128 \,S_{-1,3,-1}+128 \,S_{-1,3,1}\nonumber \\
&& +192 \,S_{1,-3,-1}+192 \,S_{1,-3,1}+96 \,S_{1,-2,-2}+96 \,S_{1,-2,2}+96 \,S_{1,2,-2}+96
   \,S_{1,2,2}+192 \,S_{1,3,-1}\nonumber \\
&& +192 \,S_{1,3,1}+256 \,S_{2,-2,-1}+224 \,S_{2,-2,1}+128 \,S_{2,-1,-2}+128 \,S_{2,-1,2}+160 \,S_{2,1,-2}\nonumber \\
&& +160 \,S_{2,1,2}+256 \,S_{2,2,-1}+288
   \,S_{2,2,1}+320 \,S_{3,-1,-1}+256 \,S_{3,-1,1}\nonumber \\
&& +320 \,S_{3,1,-1}+384 \,S_{3,1,1}~.
\ea

\medskip
\noindent
\underline{\bf L=7}

\medskip
\noindent
The argument of the harmonic sums is 
\be
S_{\cdots} \equiv S_{\cdots}(n).
\ee
\ba
\sigma_{7, 2} &=& 14 \,S_3-12 \,S_{1,2}-20 \,S_{2,1}, \\
\sigma_{7, 3} &=& 61 \,S_5-36 \,S_{1,4}-70 \,S_{2,3}-94 \,S_{3,2}-84 \,S_{4,1}+24 \,S_{1,2,2}+48 \,S_{1,3,1}\nonumber\\
&& +40 \,S_{2,1,2}+72 \,S_{2,2,1}+96 \,S_{3,1,1}~.
\ea

\bigskip
\noindent
Notice that for $s=1/2$ the general expressions
\ba
L=4, 5 &\qquad& \Sigma_L^{(s)}(N) = \sum_{n=1}^{\frac{N}{2}}\left[\sigma_L^{(s)}(n)-\sigma_L^{(s)}(0)\right],\\
L=6, 7 &\qquad& \Sigma_L^{(s)}(N) = \sum_{n=1}^{\frac{N}{2}}\,\sum_{m=1}^n \left[\sigma_L^{(s)}(m)-\sigma_L^{(m)}(0)\right],
\ea
simplify since $\sigma_L^{(1/2)}(0)$ for all the considered $L$ and up to 3 loops.

\section{Linear sum rules: Structural properties and twist-dependent formulas}
\label{sec:linear-scalar-twist}

The previous results show various remarkable structural properties. These are
\begin{enumerate}
\item The general formula for $\Sigma_L$ up to three loops takes the form 
\be
\Sigma_L(N) = \sum_{n_1=1}^{\frac{N}{2}}\,\sum_{n_2=1}^{n_1}\cdots\sum_{n_p=1}^{n_{p-1}}\,\sigma_L(n_p),
\ee
where the number of sums is $p=n-1$ for both $L=2\,n$ and $L=2\,n+1$.

\item The internal function $\sigma_L(n_p)$ can be written as a linear combination of harmonic sums with total 
transcendentality equal to $2\,\ell-1$ where $\ell$ is the loop order $\ell = 1, 2, 3$.

\item The argument of the harmonic sums is $n_p$ for odd $L$ and $2\,n_p$ for even $L$.

\item The multi-index of the harmonic sums does involve only positive indices for odd $L$.

\item The set of multi-indices is the same for all even $L$ and fixed loop order. The same is true for odd $L$ with a different 
set of indices. 
\end{enumerate}
We have extended the calculation up to $L=13$ testing the above structural properties. In all the considered cases
they hold. Also, looking at the $L$ dependence of the coefficients of the harmonic sums, we have been able to 
write down the following compact and, in our opinion, remarkable expressions.

\subsection{Odd twist}

Up to three loops, we have
\ba
\label{eq:linearodd}
\Sigma_L(N) &=&\phantom{+} 2\,(L-1)\,S_{X, 1}\,g^2 + \\
&&  + \big[(3\,L-7)\,S_{X,3}-2\,(L-1)\,S_{X, 1, 2}-4\,(L-2)\,S_{X, 2, 1}\big]\,g^4 + \nonumber\\
&&  + \big[(20\,L-79) \,S_{X,5} - 6\,(L-1) \,S_{X,1,4} -12\,(2\,L-7) \,S_{X,4,1} +\nonumber\\
&& -2\,(8\,L-21) \,S_{X,2,3} -2\,(12\,L-37) \,S_{X,3,2}
+ 4\,(L-1) \,S_{X,1,2,2} + \nonumber\\
&& + 8\,(L-2) \,S_{X,2,1,2} + 8\,(2\,L-5) \,S_{X,2,2,1} + 8\,(L-1) \,S_{X,1,3,1} + \nonumber\\
&& + 24\,(L-3) \,S_{X,3,1,1}\big]\,g^6 + \cdots ~~.\nonumber
\ea
where
\be
S_{X, \mathbf{a}}\equiv S_{X, \mathbf{a}}\left(\frac{N}{2}\right),\qquad X = \underbrace{\{0, \cdots, 0\}}_{\frac{L-3}{2}},
\ee
and a harmonic sum with trailing 0 indices is defined as 
\be
S_{0, \mathbf{a}}(N) = \sum_{n=1}^N\,S_\mathbf{a}(n).
\ee
This is the usual definition provided we define  $\mbox{sign}(0)\equiv 1$.

The above formula works in all the considered case. It also works for $L=3$ up to two loops. 
The three loop term is not covered for this initial value.

\subsection{Even twist}

Due to the larger computational complexity of the even twist case, we only present a result at the two loop level.
We define in this case
\be
\widetilde{S}_{X, \mathbf{a}}\equiv \widetilde{S}_{X, \mathbf{a}}\left(\frac{N}{2}\right),\qquad X = \underbrace{\{0, \cdots, 0\}}_{\frac{L-2}{2}}
\ee
and (notice the most inner $2\,i_p$ argument)
\be
\widetilde{S}_{\underbrace{\scriptstyle 0,\dots,0}_p,\mathbf{a}}(n) = 
\sum_{i_1=1}^n\sum_{i_2=1}^{i_1}\sum_{i=1_3}^{i_2}\cdots \sum_{i_p=1}^{i_{p-1}} S_\mathbf{a}(2\,i_p)
\ee
One finds for even $L\ge 4$
\ba
\label{eq:lineareven}
\Sigma_L^{(1/2)}(N) &=&\left[2\,L\,\widetilde{S}_{X,1} + 2\,(L-2)\,\widetilde{S}_{X,-1}\right]\,g^2 + \\
&&  + \big[
4 (3 L-8) \,\widetilde{S}_{X,-3}+12 (L-2) \,\widetilde{S}_{X, 3}-8 (L-2) \,\widetilde{S}_{X,-2,-1}-8 (L-3) \,\widetilde{S}_{X,-2,1} \nonumber\\
&& -4 (L-2) \,\widetilde{S}_{X,-1,-2}-4 (L-2) \,\widetilde{S}_{X,-1,2}-4 L \,\widetilde{S}_{X,1,-2}-4 L \,\widetilde{S}_{X,1,2} \nonumber\\
&& -8 (L-2) \,\widetilde{S}_{X,2,-1}-8 (L-1) \,\widetilde{S}_{X,2,1}
\big]\,g^4 + \cdots~. \nonumber
\ea

\section{Large $N$ check: Recovering the cusp anomalous dimension}
\label{sec:linear-checks}

An important check of the previous results Eqs.~(\ref{eq:linearodd}, \ref{eq:lineareven}) is that for large $N$ 
all the ground and excited anomalous dimensions are expected to scale logarithmically with $N$ with a coupling dependence 
reabsorbed in the physical coupling $g^2_{\rm ph}$. 

\bigskip
To check this, let $\vec{a} = (a_1, a_2, \cdots )$ and $\vec{a}_k = a_k$. If $\vec{a}_1\neq 1$ we have at large $N$
\ba
S_{X, 1, \vec{a}}(N) \sim \frac{N^p}{p\,!}\,S_{\vec{a}}(\infty)\,\log\,N,\qquad X = \underbrace{0\cdots 0}_p. 
\ea
Apart from trivial factors depending on the multiplicity of the singlet set of states, we can read the coefficient of the
logarithmic leading term by collecting all the nested harmonic sums with a leading 1 index and replacing
\be
S_{1, \vec{a}}(N)\to S_{\vec{a}}(\infty).
\ee
We now present the detailed check of the $g^2_{\rm ph}$ reshuffling for all the expressions that we have listed
in the previous sections.

\subsection{Odd twist at three loops}

In this case, we use \refeq{linearodd} and write
\be
\Sigma_L(N) \sim \frac{1}{p\,!}\left(\frac{N}{2}\right)^p\,h(g)\,\log\,N,\qquad p = \frac{L-3}{2},
\ee
where the function $h(g)$ is 
\ba
h(g) &=& 2\,(L-1)\,g^2 + \\
&&  + \big[-2\,(L-1)\,S_{2}(\infty)\big]\,g^4 + \nonumber\\
&&  + \big[-6\,(L-1) \,S_{4}(\infty) + 4\,(L-1) \,S_{2,2}(\infty) +  8\,(L-1) \,S_{3,1}(\infty)\big]\,g^6 + \cdots ~~.\nonumber 
\ea
Notice that the number of singlets is asymptotically
\be
\#\,\mbox{singlets} = \binom{\frac{N}{2}+p}{p}\sim \frac{1}{p\,!}\,\left(\frac{N}{2}\right)^p
\ee
Hence, we can divide by the multiplet dimension and write
\be
\overline{\Sigma}_L(N) \sim h(g)\,\log\,N.
\ee
Replacing the following asymptotic sums
\be
\begin{array}{ccl}
\displaystyle S_2(\infty) &=& \displaystyle  \zeta_2 = \frac{\pi^2}{6}, \\ \\
\displaystyle S_4(\infty) &=& \displaystyle   \zeta_4 = \frac{\pi^4}{90}, 
\end{array}
\qquad
\begin{array}{ccl}
\displaystyle S_{2,2}(\infty) &=& \displaystyle  \frac{7\,\pi^4}{360},\\ \\
\displaystyle S_{3,1}(\infty) &=& \displaystyle  \frac{\pi^4}{72},
\end{array}
\ee
we find
\ba
h(g) &=& 2\,(L-1)\,\left(g^2-\zeta_2\,g^4 + \frac{11\,\pi^4}{180}\,g^6 + \cdots\right) = \nonumber \\
&=& 2\,(L-1)\,g^2_{\rm ph}.
\ea

\subsection{$L=4$ at three loops}

The quantities $\sigma_{4,\ell}$ contain the following nested sums with a leading 1 (and argument $2\,n$)
\ba
\sigma_{4,1} &=& 8\,S_1+\cdots, \\
\sigma_{4,2} &=& -16\,(S_{1,2}+S_{1,-2})+\cdots, \\
\sigma_{4,3} &=& -192\,(S_{1,4}+S_{1,-4})+128\,(S_{1,3,1}+S_{1,3,-1}+S_{1,-3,1}+S_{1,-3,-1}) +\nonumber\\
&& +  64\,
(S_{1,2,2}+S_{1,2,-2}+S_{1,-2,2}+S_{1,-2,-2})+\cdots~.
\ea
The required asymptotic values are
\ba
\left. S_{2}+S_{-2} \right|_\infty &=& \frac{\pi^2}{12}, \\
\left. S_{4}+S_{-4} \right|_\infty &=& \frac{\pi^4}{720}, \\
\left. S_{3,1}+S_{3,-1}+S_{-3,1}+S_{-3,-1} \right|_\infty &=& \frac{\pi^4}{288}, \\
\left. S_{2,2}+S_{2,-2}+S_{-2,2}+S_{-2,-2} \right|_\infty &=& \frac{7\,\pi^4}{1440}.
\ea
Collecting, we find
\ba
\sigma_4 &\sim& 8\,\log\,N\,\left(g^2-\zeta_2\,g^4+\frac{11\,\pi^4}{180}\,g^6 + \cdots\right) = \nonumber \\
 &=& 8\,\log\,N\,g^2_{\rm ph}~.
\ea

\subsection{$L=6$ at three loops}

The quantities $\sigma_{6,\ell}$ contain the following nested sums with a leading 1 (and argument $2\,n$)
\ba
\sigma_{6,1} &=& 12\,S_1+\cdots, \\
\sigma_{6,2} &=& -24\,(S_{1,2}+S_{1,-2})+\cdots, \\
\sigma_{6,3} &=& -288\,(S_{1,4}+S_{1,-4})+192\,(S_{1,3,1}+S_{1,3,-1}+S_{1,-3,1}+S_{1,-3,-1}) +\nonumber\\
&& +  96\,
(S_{1,2,2}+S_{1,2,-2}+S_{1,-2,2}+S_{1,-2,-2})+\cdots~.
\ea
With the previous asymptotic values, we find
\ba
\sigma_6 &\sim& 12\,\log\,N\,\left(g^2-\zeta_2\,g^4+\frac{11\,\pi^4}{180}\,g^6 + \cdots\right) = \nonumber \\
&=& 12\,\log\,N\,g^2_{\rm ph}~.
\ea

\subsection{General even $L$ at two loops}

The asymptotic values given for the cases $L=4, 6$ are sufficient to check the two loop general even $L$ case.

\section{Linear sum rules: Subleading corrections at large $N$}
\label{sec:linear-subleading}

We can take our master formula for odd twist \refeq{linearodd} and expand $\overline{\Sigma}_L(N)$ at large $N$ computing the subleading
part analogous to $f_{\rm sl}$ defined in \refeq{generalized1}. The expansion requires a treatment of multiple sums with 
various trailing zeroes of the form 
\be
S_{\underbrace{\scriptstyle 0, \dots, 0}_p, \mathbf{X}}.
\ee
They can be treated with the nice results reported in App.~(\ref{app:zero}). It is clear that some structure immediately arises.
For instance, let us consider the twist 5 case where $p=1$. The leading terms in $\Sigma$ , proportional to $N$, arise
from sums of the form $S_{0, 1, a, \mathbf{X}}(n)$ where $n=N/2$ and $a>1$.
From \refeq{zeroone} of App.~(\ref{app:zero}), we get
\ba
S_{0, 1, a, \mathbf{X}}(n) &=& (n+1)\,S_{1, a, \mathbf{X}}(n)-S_{0, a, \mathbf{X}}(n) = \\
&=& (n+1)\left[S_{1, a, \mathbf{X}}(n)-S_{a, \mathbf{X}}(n)\right]+S_{a-1, \mathbf{X}}(n).
\ea
For large $n$, we need only the first bracket whose expansion contains the terms
\be
S_{1, a, \mathbf{X}}(n)-S_{a, \mathbf{X}}(n) = (\log\,n + \gamma_{\rm E}-1)\,\zeta_{a, \mathbf{X}} + \dots, 
\ee
where dots denote other constant terms. The first part combines to give the physical coupling which thus must appear in the 
combination
\be
(\log\,n + \gamma_{\rm E}-1)\,g^2_{\rm ph}.
\ee
The prefactor is easily generalized to a generic odd twist, {\em i.e.} $p>1$. The leading term of  $S_{\underbrace{\scriptstyle 0,...,0}_{p},1,X}(n)$
for large $n$ replaces it by the general form 
\begin{equation}
\log\,n + \gamma_{\rm E}-1 \ \to\ \frac{1}{p!} \left[a_p(\log n+\gamma_{\rm E})-b_p\right],
\end{equation}
where 
\ba
a_p &=& a_{p-1} = 1, \\
b_p &=& b_{p-1}-\frac{1}{p},\qquad b_1=1,\ \la \ b_p = S_1(p).
\ea
The remaining subleading pieces can be worked out in a similar way. 
The final result is 
\ba
\label{eq:newsubl}
\overline\Sigma_L(N) &=& 2\,(L-1)\,\left[\log\,\frac{N}{2}+\gamma_{\rm E}-S_1\left(\frac{L-3}{2}\right)\right]\,g^2_{\rm ph} + \\
&& -(3\,L-7)\,\zeta_3\,g^4 + \nonumber\\
&& + \left[\frac{L-2}{3}\,\pi^2\,\zeta_3 + (10\,L-31)\,\zeta_5\right]\,g^6 + \cdots~.\nonumber
\ea
\refeq{newsubl} is the generalization of the recent  result \refeq{generalized2}. The structure is quite similar, although 
\refeq{newsubl} involves the sum over the singlet anomalous dimensions !

\section{Higher order sum rules: The quadratic case}
\label{sec:quadratic}

Let us define higher order sum rules by considering sums of powers of the individual anomalous dimensions.
In particular, we focus on the quadratic sum
\be
{\cal Q}_L^{(s)}(N) = \sum_{k\in \ \rm singlets}\left[\gamma_{L, k}^{(s)}(N)\right]^2.
\ee
Remarkably, we find simple sum rules also for these higher order sums. We illustrate this in the special cases
$L=4,5$ for $s=1/2$.
We find again the general representation (valid for $L=4,5$)
\be
{\cal Q}_L(N) = \sum_{n=1}^{N/2}\,\sum_{\ell\ge 1} g^{2\,\ell+2}\,q_{L, \ell}(n),
\ee
where 

\medskip
\noindent
\underline{\bf L=4}

\medskip
\noindent
The argument of the harmonic sums is 
\be
S_{\cdots} \equiv S_{\cdots}(2\,n).
\ee

\ba
q_{4, 1} &=& -48 \,S_{-2}-128 \,S_2+64 \,S_{-1,-1}+32 \,S_{-1,1}+64 \,S_{1,-1}+128 \,S_{1,1}, \\
q_{4, 2} &=& -384 \,S_{-4}-832 \,S_4+576 \,S_{-3,-1}+256 \,S_{-3,1}+704 \,S_{-2,-2}+384 \,S_{-2,2}\nonumber\\
&& +576 \,S_{-1,-3}+384 \,S_{-1,3}+768 \,S_{1,-3}+1152 \,S_{1,3}+768 \,S_{2,-2}+1280 \,S_{2,2}+576
   \,S_{3,-1}\nonumber\\
&& +1024 \,S_{3,1}-256 \,S_{-2,-1,-1}-512 \,S_{-2,-1,1}-256 \,S_{-2,1,-1}-128 \,S_{-2,1,1}-256 \,S_{-1,-2,-1}\nonumber\\
&& -384 \,S_{-1,-2,1}-256 \,S_{-1,-1,-2}-256 \,S_{-1,-1,2}-128
   \,S_{-1,1,-2}-128 \,S_{-1,1,2}-256 \,S_{-1,2,-1}\nonumber\\
&& -128 \,S_{-1,2,1}-512 \,S_{1,-2,-1}-256 \,S_{1,-2,1}-256 \,S_{1,-1,-2}-256 \,S_{1,-1,2}\nonumber\\
&& -512 \,S_{1,1,-2}-512 \,S_{1,1,2}-512
   \,S_{1,2,-1}-768 \,S_{1,2,1}-512 \,S_{2,-1,-1}-256 \,S_{2,-1,1}\nonumber\\
&& -512 \,S_{2,1,-1}-896 \,S_{2,1,1}.
\ea

\medskip
\noindent
\underline{\bf L=5}

\medskip
\noindent
The argument of the harmonic sums is 
\be
S_{\cdots} \equiv S_{\cdots}(n).
\ee

\ba
q_{5, 1} &=& 128 \,S_{1,1}-88 \,S_2, \\
q_{5, 2} &=& -212 \,S_4+384 \,S_{1,3}+432 \,S_{2,2}+352 \,S_{3,1}-256 \,S_{1,1,2}\nonumber\\
&& -384 \,S_{1,2,1}-448 \,S_{2,1,1}, \\
q_{5, 3} &=& -702 \,S_6+1616 \,S_{1,5}+2112 \,S_{2,4}+2352 \,S_{3,3}+2072 \,S_{4,2}+1408 \,S_{5,1}\nonumber\\
&& -1536 \,S_{1,1,4}-2080 \,S_{1,2,3}-2464 \,S_{1,3,2}-2304 \,S_{1,4,1}-2336 \,S_{2,1,3}\nonumber\\
&& -2784
   \,S_{2,2,2}-2768 \,S_{2,3,1}-2912 \,S_{3,1,2}-2912 \,S_{3,2,1}-2544 \,S_{4,1,1}+768 \,S_{1,1,2,2}\nonumber\\
&& +1536 \,S_{1,1,3,1}+1152 \,S_{1,2,1,2}+1920 \,S_{1,2,2,1}+2304 \,S_{1,3,1,1}+1344
   \,S_{2,1,1,2}\nonumber\\
&& +2112 \,S_{2,1,2,1}+2496 \,S_{2,2,1,1}+2688 \,S_{3,1,1,1}.
\ea

\section{Quadratic sum rules: Structural properties and twist dependent formulas}
\label{sec:quadratic-twist}

In complete analogy with the linear sum rules, we can observe basically the same structural properties also in the case
of the quadratic sum rules. In particular
the general formula for ${\cal Q}_L$ seems to be 
\be
{\cal Q}_L(N) = \sum_{n_1=1}^{\frac{N}{2}}\,\sum_{n_2=1}^{n_1}\cdots\sum_{n_p=1}^{n_{p-1}}\,q_L(n_p),
\ee
where the number of sums is again $p=n-1$ for both $L=2\,n$ and $L=2\,n+1$, but now the total 
transcendentality of the sums in $q_L$ is equal to $2\,\ell$ where $\ell$ is the loop order $\ell = 1, 2, 3$.

Also in this case, we have extended the calculation up to $L=13$. Now, the $L$ dependence of the harmonic sums coefficients is quadratic
instead of linear and we can write down the following compact expressions.

\subsection{Odd twist}

We have, at two loops, 
\ba
{\cal Q}_L(N) &=& \big[8\, (L-1)^2\, S_{X, 1,1}-4 \left(2 L^2-7 L+7\right)\, S_{X,2}\big]\,g^4 + \\
&& + \big[
-16 \, (L-1)^2\,S_{X, 1,1,2}+12 \,(3 L-7)  (L-1)\,S_{X, 1,3} \nonumber\\
&& -32 \,(L-2)  (L-1) \, S_{X, 1,2,1}-4 \,
\left(12 L^2-74 L+123\right)\, S_{X, 4}\nonumber\\
&& +4\, \left(13 L^2-58 L+73\right)\, S_{X, 2,2}+8\, \left(7 L^2-38
   L+59\right) \, S_{X, 3,1}\nonumber\\
&& -16 \, \left(3 L^2-12 L+13\right) \,  S_{X, 2,1,1}
\big]\,g^6 + \cdots~. \nonumber
\ea

where, again, 
\be
S_{ X, \mathbf{a}}\equiv S_{X, \mathbf{a}}\left(\frac{N}{2}\right),\qquad X = \underbrace{\{0, \cdots, 0\}}_{\frac{L-3}{2}}
\ee

The cusp anomaly check is clearly passed by the combination 
\be
8\, (L-1)^2\, S_{X, 1,1} \,g^4  -16 \, (L-1)^2\,S_{X, 1,1,2} \, g^6 .
\ee

\subsection{Even twist}

Due to the larger computational complexity of the even twist case, we only present a one-loop result.
We define in this case
\be
\widetilde{S}_{X, \mathbf{a}}\equiv \widetilde{S}_{X, \mathbf{a}}\left(\frac{N}{2}\right),\qquad X = \underbrace{\{0, \cdots, 0\}}_{\frac{L-2}{2}}
\ee
and (notice the most inner $2\,i_p$ argument)
\be
\widetilde{S}_{\underbrace{\scriptstyle 0,\dots,0}_p,\mathbf{a}}(n) = 
\sum_{i_1=1}^n\sum_{i_2=1}^{i_1}\sum_{i=1_3}^{i_2}\cdots \sum_{i_p=1}^{i_{p-1}} S_\mathbf{a}(2\,i_p)
\ee
One finds for even $L\ge 4$
\ba
{\cal Q}_L(N) &=&\big[
-8\, \left(2 L^2-5 L+4\right)\, S_{X,2}
-8\, (2 L-5) \, (L-2)\,S_{X,-2} \nonumber\\
&& 
+8\, L^2\, S_{X,1,1}
+8\, L\,(L-2)\, S_{X,1,-1} \nonumber\\
&& + 8\, (L-2)^2\,S_{X,-1,1} 
+8\, L\, (L-2)\, S_{X,-1,-1} 
\big]\,g^4 + \cdots
\ea

\section{Large $N$ check for the quadratic sum rules}
\label{sec:quadratic-checks}

Here we report the cusp anomalous dimension check for the various formulas computing quadratic sum rules at $L=4$ (two loops) and $L=5$ (three loops).

\subsection{L=4 at two loops}

The squared logarithmic terms are
\ba
q_{4, 1} &=& 128 S_{1,1} + \cdots, \\
q_{4, 2} &=& -512 S_{1,1,-2}-512 S_{1,1,2} + \cdots .
\ea
Hence, at two loops
\ba
q_4 &=& 64\,\log^2\,N\,(g^4-2\,\zeta_2\,g^6+\cdots) = \nonumber \\
&=&  64\,\log^2\,N\,(g^2-\zeta_2\,g^4+\cdots)^2~ = \nonumber \\
&=&  64\,\log^2\,N\,g^4_{\rm ph}.
\ea

\subsection{L=5 at three loops}

The squared logarithmic terms are
\ba
q_{5, 1} &=& 128 S_{1,1} + \cdots, \\
q_{5, 2} &=& -256 S_{1,1,2} + \cdots,\\
q_{5, 3} &=& -1536 S_{1,1,4}+768 S_{1,1,2,2} +1536 S_{1,1,3,1}+\cdots .
\ea

Hence, at three loops
\ba
q_5 &=& 64\,\log^2\,N\,(g^4-2\,\zeta_2\,g^6+\frac{3\,\pi^4}{20}\,g^8+\cdots) = \nonumber \\
&=&  64\,\log^2\,N\,(g^2-\zeta_2\,g^4+\frac{11\,\pi^4}{180}\,g^6+\cdots)^2~ \nonumber \\
&=&  64\,\log^2\,N\,g^4_{\rm ph}~.
\ea

\section{One loop cubic sum rule}
\label{sec:cubic}

It is clear that it is possible to derive sum rules at arbitrary high order. The determination of the explicit formulas
is a matter of computational effort. Here, we just give, as an example, the cubic sum rule
\be
{\cal C}_L^{(s)}(N) = \sum_{k\in \ \rm singlets}\left[\gamma_{L, k}^{(s)}(N)\right]^3,
\ee
for the scalar sector $s=1/2$ and odd twist $L$ at one-loop. We have tested it again up to $L=13$. It reads
\ba
{\cal C}_L(N) &=& \big[
48\, (L-1)^3\, S_{X, 1,1,1}-24\, (L-1)\,\left(2 L^2-7 L+7\right)\, S_{X, 1,2} \nonumber\\
&& -48\, (L-2)\, \left(L^2-4 L+7\right)\, S_{X, 2,1} \nonumber\\
&& +8\, \left(6 L^3-45 L^2+124 L-121\right)\, S_{X, 3}\big]\,g^6 + \cdots~.
\ea
where, again, 
\be
S_{ X, \mathbf{a}}\equiv S_{X, \mathbf{a}}\left(\frac{N}{2}\right),\qquad X = \underbrace{\{0, \cdots, 0\}}_{\frac{L-3}{2}}
\ee

\section{Conclusions}
\label{sec:conclusions}

In summary, we have shown that it is useful to define higher-order sum rules for the anomalous dimensions of 
singlet unpaired twist-operators in ${\cal N}=4$ SYM for arbitrarily high twist. Of course, these combinations contain less information than the separate
anomalous dimensions. However, on the other hand, they admit multi-loop closed expressions in terms of the usual nested harmonic sums 
which are ubiquitous in this context. These expressions provide interesting handles for analytical calculations in higher twist.

The basic technical hint behind the sum rules is that they are related to restricted traces of powers of the dilatation operator. As such, they are quite simpler
objects than the separate anomalous dimensions. As an analogy, it is typically simpler to focus on the coefficient of a polynomial instead of looking at its 
explicit and complicated roots.

Our multi-loop results have been obtained in the $\mathfrak{sl}(2)$ sector where the long-range Bethe equations are particularly simple.
At one-loop, analogous results have been presented for the other basic sectors describing purely fermionic or gauge operators. 
It should be possible to extend the analysis to these cases at higher loops by working out the perturbative expansion of the 
relevant higher rank long-range equations.

It remains to be understood if the closed expressions we found for the sum rules are just a curiosity or a manifestation of deeper
properties. In this respect, their interpretation in the light 
of AdS/CFT duality would certainly be a very interesting issue. 
Indeed, in this context, the linear sum rules compute sums of energies of dual string configurations with a fixed number of spikes,
but different values of the internal degrees of freedom associated with the band of states (for twist $>2$)~\cite{Belitsky:2003ys}. 
Hence, on the string side, the proposed sum rule suggest to investigate the properties of energies after a sort of averaging over these kinematical features. The powerful analysis 
in~\cite{Kruczenski:2008bs} could prove to be useful in this respect.

\acknowledgments

We thank M. Staudacher, V. Forini, S. Zieme, G. Marchesini and Y. L. Dokshitzer for useful discussions and comments.
M.~B. warmly thanks AEI Potsdam-Golm  for very kind hospitality while working on parts of this project. 

\appendix

\section{Symmetric polynomials and sums of powers}
\label{app:symmetric}

Given a finite set of complex numbers $x_1, \dots, x_n\in \mathbb{C}$, the symmetric polynomials $\Pi_k(x_1, \dots, x_n)$
are defined as
\be
\prod_{i=1}^n (x-x_i) = \sum_{k=0}^n (-1)^k\,\Pi_k(x_1, \dots, x_n)\,x^k.
\ee
Hence, 
\ba
\Pi_0(x_1, \dots, x_n) &=& 1, \\ 
\Pi_1(x_1, \dots, x_n) &=& x_1+\cdots x_n, \\ 
\Pi_2(x_1, \dots, x_n) &=& \sum_{1\le i < j \le n} x_i\,x_j, \\
&\cdots& \\
\Pi_n(x_1, \dots, x_n) &=& \prod_{i=1}^n x_i.
\ea
The relation with the symmetric sums of powers
\be
S_k = \sum_{i=1}^k x_i^k,
\ee
is given by the generating function relation
\be
\sum_{k=0}^\infty \Pi_k\,t^k = \exp\left( \sum_{k=1}^\infty (-1)^{k+1}\,\frac{S_k}{k}\,t^k\right),
\ee
or by the Newton-Girard recursion 
\be
(-1)^m\,m\,\Pi_m + \sum_{k=1}^m\,(-1)^{k+m}\,\Pi_{m-k}\,S_k = 0.
\ee
The resulting map $(\Pi_1, \dots, \Pi_n)\leftrightarrow (S_1, \dots, S_n)$ is birational, actually polynomial, and begins with 
\be
\begin{array}{ccl}
S_1 &=& \Pi_1, \\
S_2 &=& \Pi_1^2-2\,\Pi_2, \\
S_3 &=& \Pi_1^3-3\,\Pi_1\,\Pi_2+3\,\Pi_3,
\end{array}
\qquad
\begin{array}{ccl}
\Pi_1 &=& S_1, \\
\Pi_2 &=& \frac{1}{2}\left(S_1^2-S_2\right), \\
\Pi_3 &=& \frac{1}{6}\left(S_1^3-3\,S_1\,S_2+2\,S_3\right).
\end{array}
\ee

\section{Rationality proof}
\label{app:rationality}

\begin{theorem} Let $R(x)$ be a rational function of $x$ over $\mathbb{Q}$, i.e. the ratio of two polynomials in $\mathbb{Q}[x]$.
For any polynomial with rational coefficients $P(x)\in \mathbb{Q}[x]$ one has
\be
\sum_{x\in\mathbb{C} : \ P(x)=0} R(x) \in \mathbb{Q}~.
\ee
\end{theorem}
{\bf Proof:} 

Let $d = \deg\,P$. For any root of $P(x)=0$ we can write the identity
\be
R(x) = \sum_{n=0}^{d-1} c_n\,x^n,
\ee
with suitable coefficients $\{c_n\}$ which are rational functions of the coefficients appearing in $R$. They are the same for all roots of $P$. 
As is well known (see App.~(\ref{app:symmetric})), the sums 
\be
S_n = \sum_{x\in\mathbb{C}\ : \ P(x)=0} x^n,
\ee
are all fully determined as rational functions of the coefficients of $P(x)$. Thus they are rational and therefore
\be
\sum_{x\in\mathbb{C} : \ P(x)=0} R(x) =  \sum_{n=0}^{d-1} c_n\,S_n \in \mathbb{Q}.
\ee

\hfill $\Diamond$

This theorem can be greatly extended by replacing $x$ by a finite set of variables and $P(x)=0$ by a system of polynomial 
equations with rational coefficients. The proof is the same as in the one dimensional case after reduction by Gr\"obner basis methods~\cite{Grob}.

\bigskip
In order to illustrate the above (constructive) theorem in the univariate case, let us consider in details as a simple example the derivation of 
\be
\sum_{x^{6}+x+1=0}\frac{1}{x} = -1.
\ee
We start from the relation 
\be
\frac{1}{x} = -1 -x^5,\qquad \mbox{iff}\qquad  x^6+x+1=0.
\ee
Hence
\be
\sum_{x^6+x+1=0}\frac{1}{x} = -6-\sum_{x^6+x+1=0}x^5.
\ee
From the results of App.~(\ref{app:symmetric}) we compute for $P(x) = x^6+x+1$ the explicit sums of powers
\be
S_1 = S_2 = S_3 = S_4 = 0,\qquad S_5 = -5.
\ee
Hence, we have proved that
\be
\sum_{x^{6}+x+1=0}\frac{1}{x} = -6-S_5 =-1.
\ee

\section{Harmonic sums with trailing 0 indices and positive indices}
\label{app:zero}

A leading 0 index means
\be
S_{0, \bf{X}}(N) = \sum_{n=1}^N S_{\bf{X}}(n).
\ee
Clearly, we have
\ba
S_{0}(N) &=& N, \\
S_{0,0}(N) &=& \frac{N(N+1)}{2\,!},
\ea
and the general formula
\be
S_{\underbrace{\scriptstyle 0, \dots, 0}_p}(N) = \binom{N+p-1}{p}.
\ee
The first non trivial result is 
\begin{theorem}
For any $a\ge 1$, we have 
\be
\label{eq:zeroone}
S_{0, a, \bf{X}}(N) = (N+1)\,S_{a, \bf{X}}-S_{a-1, \bf{X}},
\ee
\end{theorem}
{\bf Proof:}

We can prove the theorem by induction on $N$. Taking the difference of the equation between $N+1$ and $N$ we find 
$$
S_{0, a, \bf{X}}(N+1) - (N+2)\,S_{a, \bf{X}}(N+1)+S_{a-1, \bf{X}}(N+1)-S_{0, a, \bf{X}}(N) + (N+1)\,S_{a, \bf{X}}(N)-S_{a-1, \bf{X}}(N) = 
$$
$$
= S_{a, \bf{X}}(N+1) - (N+2)\,S_{a, \bf{X}}(N+1)+\frac{1}{(N+1)^{a-1}} S_{\bf{X}}(N+1) + (N+1)\,S_{a, \bf{X}}(N) = 
$$
$$
= S_{a, \bf{X}}(N+1) - S_{a, \bf{X}}(N) -\frac{N+2}{(N+1)^a}\,S_{\bf{X}}(N+1)+\frac{1}{(N+1)^{a-1}} S_{\bf{X}}(N+1) = 
$$
$$
= S_{a, \bf{X}}(N+1) - S_{a, \bf{X}}(N) -\frac{1}{(N+1)^a}\,S_{\bf{X}}(N+1) = 0.
$$
\hfill$\Diamond$

\medskip
\noindent
The generalization is
\begin{theorem}
For any $a\ge 1$, we have 
\be
\label{eq:thesis}
S_{\underbrace{\scriptstyle 0, \dots, 0}_p, a, \bf{X}}(N) = \frac{1}{p}\left[(N+p)\,
S_{\underbrace{\scriptstyle 0, \dots, 0}_{p-1}, a, \bf{X}}-S_{\underbrace{\scriptstyle 0, \dots, 0}_{p-1}, a-1, \bf{X}}\right].
\ee
\end{theorem}
{\bf Proof:} it follows from induction over $p$.
Let us assume that \refeq{thesis} is true for $p$ for all $N$, then we can prove that it is true for $p+1$ by induction over $N$. 
Following similar steps as in the proof of Theorem C.1 we find
\begin{displaymath}
S_{\underbrace{\scriptstyle 0, \dots ,0}_{p+1}, a, X}(N+1)-\frac{1}{p+1}\left[(N+p+2)S_{\underbrace{\scriptstyle 0, \dots ,0}_{p}, a, X}(N+1)
-S_{\underbrace{\scriptstyle 0, \dots ,0}_{p}, a-1, X}(N+1) \right]
\end{displaymath}
\begin{displaymath}
-S_{\underbrace{\scriptstyle 0, \dots ,0}_{p+1}, a, X}(N)+\frac{1}{p+1}\left[(N+p+1)S_{\underbrace{\scriptstyle 0, \dots ,0}_{p}, a, X}(N)
-S_{\underbrace{\scriptstyle 0, \dots ,0}_{p}, a-1, X}(N) \right]
\end{displaymath}
\begin{displaymath}
=S_{\underbrace{\scriptstyle 0, \dots ,0}_{p}, a, X}(N+1)-\frac{N+p+1}{p+1}S_{\underbrace{\scriptstyle 0, \dots ,0}_{p-1}, a, X}(N+1)
-\frac{1}{p+1}S_{\underbrace{\scriptstyle 0, \dots ,0}_{p}, a, X}(N+1)
\end{displaymath}
\begin{displaymath}
+\frac{1}{p+1}S_{\underbrace{\scriptstyle 0, \dots ,0}_{p-1}, a-1, X}(N+1)\quad = 
\quad\frac{p}{p+1}S_{\underbrace{\scriptstyle 0, \dots ,0}_{p}, a, X}(N+1)
\end{displaymath}
\begin{displaymath}
-\frac{1}{p+1}\left[(N+p+1)S_{\underbrace{\scriptstyle 0, \dots ,0}_{p-1}, a, X}(N+1)
-S_{\underbrace{\scriptstyle 0, \dots ,0}_{p-1}, a-1, X}(N+1)\right]=0
\end{displaymath}

\hfill$\Diamond$

\newpage

\vskip 2cm
\FIGURE{\epsfig{file=spectrum.L3.eps, width=16cm}
        \bigskip\bigskip
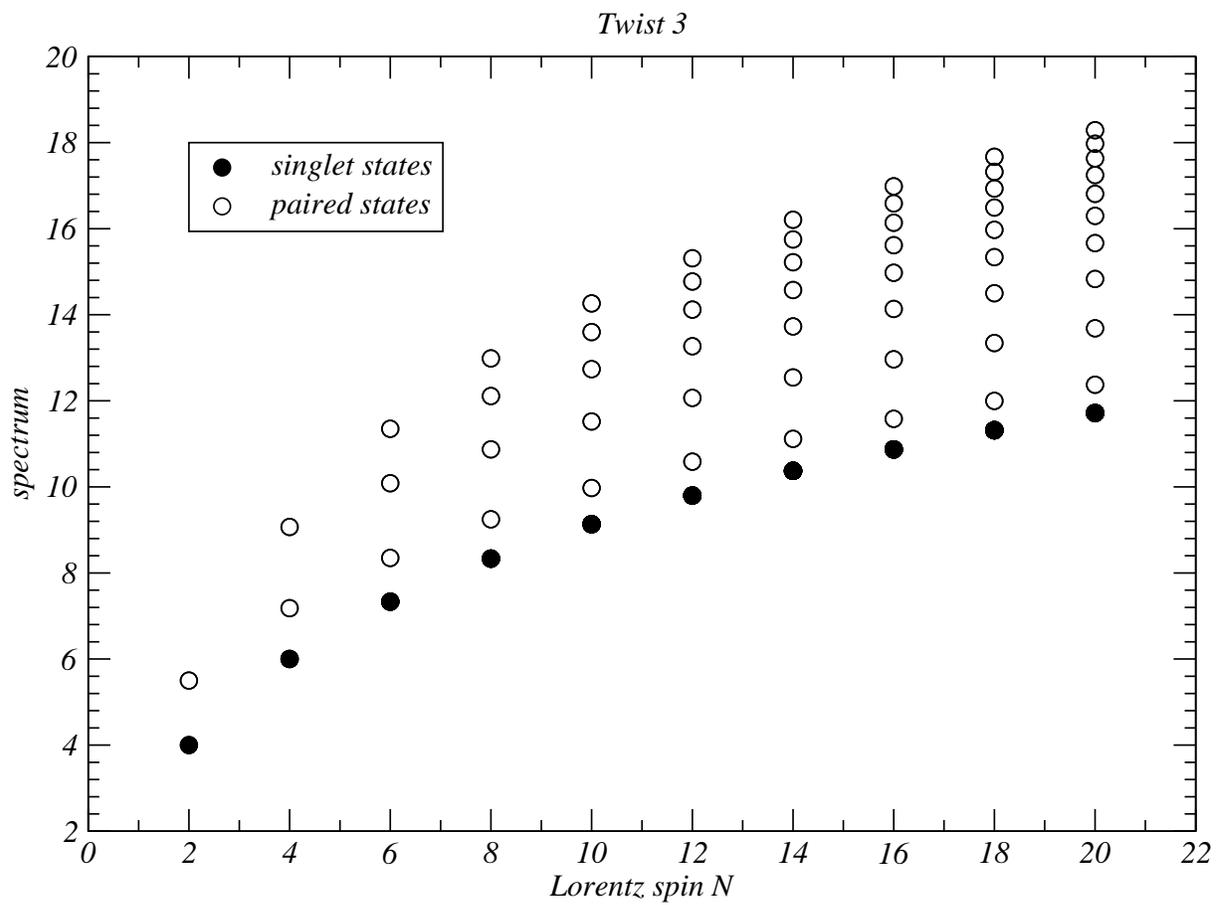
        \caption{Full spectrum at twist $L=3$.}
        \label{fig:spectrum3}}

\newpage
\vskip 2cm
\FIGURE{\epsfig{file=spectrum.L4.eps, width=15cm}
        \bigskip\bigskip
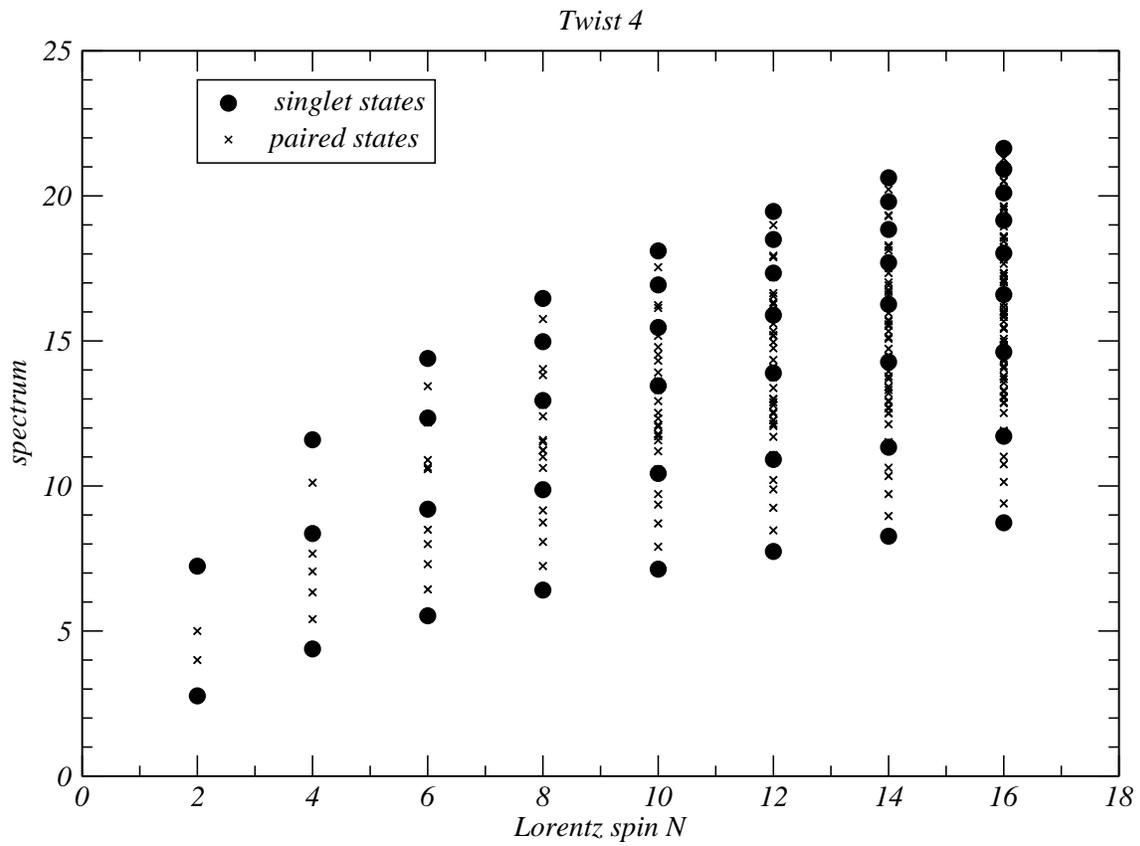
        \caption{Full spectrum at twist $L=4$.}
        \label{fig:spectrum4}}


\begin{thebibliography}{99}

%\cite{Beisert:2005fw}
\bibitem{Beisert:2005fw}
  N.~Beisert and M.~Staudacher,
  {\em Long-range PSU(2,2|4) Bethe ansaetze for gauge theory and strings}, 
  Nucl.\ Phys.\  B {\bf 727}, 1 (2005)
  [arXiv:hep-th/0504190].
  %%CITATION = NUPHA,B727,1;%%

%\cite{Ambjorn:2005wa}
\bibitem{Ambjorn:2005wa}
  See for instance, 
  J.~Ambjorn, R.~A.~Janik and C.~Kristjansen,
  {\em Wrapping interactions and a new source of corrections to the spin-chain string duality}, 
  Nucl.\ Phys.\  B {\bf 736}, 288 (2006)
  [arXiv:hep-th/0510171].
  %%CITATION = NUPHA,B736,288;%%

%\cite{Ferretti:2004ba}
\bibitem{Ferretti:2004ba}
  G.~Ferretti, R.~Heise and K.~Zarembo,
  {\em New integrable structures in large-$N$ QCD}, 
  Phys.\ Rev.\  D {\bf 70}, 074024 (2004)
  [arXiv:hep-th/0404187].
  %%CITATION = PHRVA,D70,074024;%%

%\cite{Beisert:2004fv}
\bibitem{Beisert:2004fv}
  N.~Beisert, G.~Ferretti, R.~Heise and K.~Zarembo,
  {\em One-loop QCD spin chain and its spectrum}, 
  Nucl.\ Phys.\  B {\bf 717}, 137 (2005)
  [arXiv:hep-th/0412029].
  %%CITATION = NUPHA,B717,137;%%

%\cite{Rej:2007vm}
\bibitem{Rej:2007vm}
  A.~Rej, M.~Staudacher and S.~Zieme,
  {\em Nesting and dressing}, 
  J.\ Stat.\ Mech.\  {\bf 0708}, P08006 (2007)
  [arXiv:hep-th/0702151].
  %%CITATION = JSTAT,0708,P08006;%%


%\cite{Beccaria:2007gu}
\bibitem{Beccaria:2007gu}
  M.~Beccaria and V.~Forini,
  {\em Anomalous dimensions of finite size field strength operators in ${\cal N}=4$ SYM}, 
  JHEP {\bf 0711}, 031 (2007)
  [arXiv:0710.0217 [hep-th]].
  %%CITATION = JHEPA,0711,031;%%

%\cite{Belitsky:2004cz}
\bibitem{Belitsky:2004cz}
  A.~V.~Belitsky, V.~M.~Braun, A.~S.~Gorsky and G.~P.~Korchemsky,
  {\em Integrability in QCD and beyond},
  To be published in the memorial volume {\em From Fields to Strings: Circumnavigating Theoretical Physics}, World Scientific, 2004. 
  Dedicated to the memory of Ian Kogan. 
  Int.\ J.\ Mod.\ Phys.\  A {\bf 19}, 4715 (2004)
  [arXiv:hep-th/0407232].
  %%CITATION = IMPAE,A19,4715;%%

%\cite{Beisert:2003jj}
\bibitem{Beisert:2003jj}
  N.~Beisert,
  {\em The complete one-loop dilatation operator of ${\cal N} = 4$ super Yang-Mills theory}, 
  Nucl.\ Phys.\  B {\bf 676}, 3 (2004)
  [arXiv:hep-th/0307015].
  %%CITATION = NUPHA,B676,3;%%

%\cite{Braun:2003rp}
\bibitem{Braun:2003rp}
  V.~M.~Braun, G.~P.~Korchemsky and D.~Mueller,
  {\em The uses of conformal symmetry in QCD}, 
  Prog.\ Part.\ Nucl.\ Phys.\  {\bf 51}, 311 (2003)
  [arXiv:hep-ph/0306057].
  %%CITATION = PPNPD,51,311;%%

%\cite{Belitsky:2003sh}
\bibitem{Belitsky:2003sh}
  A.~V.~Belitsky, S.~E.~Derkachov, G.~P.~Korchemsky and A.~N.~Manashov,
  {\em Superconformal operators in ${\cal N} = 4$ super-Yang-Mills theory}, 
  Phys.\ Rev.\  D {\bf 70}, 045021 (2004)
  [arXiv:hep-th/0311104].
  %%CITATION = PHRVA,D70,045021;%%

%\cite{Belitsky:2005gr}
\bibitem{Belitsky:2005gr}
  A.~V.~Belitsky, S.~E.~Derkachov, G.~P.~Korchemsky and A.~N.~Manashov,
  {\em Superconformal operators in Yang-Mills theories on the light-cone}, 
  Nucl.\ Phys.\  B {\bf 722}, 191 (2005)
  [arXiv:hep-th/0503137].
  %%CITATION = NUPHA,B722,191;%%

\bibitem{klov}
%\cite{Kotikov:2001sc}
%\bibitem{Kotikov:2001sc}
  A.~V.~Kotikov and L.~N.~Lipatov,
  {\em DGLAP and BFKL evolution equations in the ${\cal N} = 4$ supersymmetric gauge theory},
  arXiv:hep-ph/0112346.
  %%CITATION = HEP-PH/0112346;%%

%\cite{Kotikov:2002ab}
%\bibitem{Kotikov:2002ab}
  A.~V.~Kotikov and L.~N.~Lipatov,
  {\em DGLAP and BFKL equations in the ${\cal N} = 4$ supersymmetric gauge theory},
  Nucl.\ Phys.\  B {\bf 661}, 19 (2003)
  [Erratum-ibid.\  B {\bf 685}, 405 (2004)]
  [arXiv:hep-ph/0208220].
  %%CITATION = NUPHA,B661,19;%%

%\cite{Kotikov:2004er}
%\bibitem{Kotikov:2004er}
  A.~V.~Kotikov, L.~N.~Lipatov, A.~I.~Onishchenko and V.~N.~Velizhanin,
  {\em Three-loop universal anomalous dimension of the Wilson operators in ${\cal N}=  4$  SUSY Yang-Mills model},
  Phys.\ Lett.\  B {\bf 595}, 521 (2004)
  [Erratum-ibid.\  B {\bf 632}, 754 (2006)]
  [arXiv:hep-th/0404092].
  %%CITATION = PHLTA,B595,521;%%

%\cite{Beisert:2004di}
\bibitem{Beisert:2004di}
  N.~Beisert, M.~Bianchi, J.~F.~Morales and H.~Samtleben,
  {\em Higher spin symmetry and ${\cal N} = 4$ SYM}, 
  JHEP {\bf 0407}, 058 (2004)
  [arXiv:hep-th/0405057].
  %%CITATION = JHEPA,0407,058;%%

%\cite{Beccaria:2007cn}
\bibitem{Beccaria:2007cn}
  M.~Beccaria,
  {\em Anomalous dimensions at twist-3 in the $\mathfrak{sl}(2)$ sector of ${\cal N} = 4$ SYM}, 
  JHEP {\bf 0706}, 044 (2007)
  [arXiv:0704.3570 [hep-th]].
  %%CITATION = JHEPA,0706,044;%%

%\cite{Kotikov:2007cy}
\bibitem{Kotikov:2007cy}
  A.~V.~Kotikov, L.~N.~Lipatov, A.~Rej, M.~Staudacher and V.~N.~Velizhanin,
  {\em Dressing and Wrapping}, 
  J.\ Stat.\ Mech.\  {\bf 0710}, P10003 (2007)
  [arXiv:0704.3586 [hep-th]].
  %%CITATION = JSTAT,0710,P10003;%%

%\cite{Beccaria:2007bb}
\bibitem{Beccaria:2007bb}
  M.~Beccaria, Yu.~L.~Dokshitzer and G.~Marchesini,
  {\em Twist 3 of the $\mathfrak{sl}(2)$ sector of ${\cal N}=4$ SYM and reciprocity respecting
    evolution}, 
  Phys.\ Lett.\  B {\bf 652}, 194 (2007)
  [arXiv:0705.2639 [hep-th]].
  %%CITATION = PHLTA,B652,194;%%

%\cite{Beccaria:2007vh}
\bibitem{Beccaria:2007vh}
  M.~Beccaria,
  {\em Universality of three gaugino anomalous dimensions in ${\cal N} = 4$ SYM}, 
  JHEP {\bf 0706}, 054 (2007)
  [arXiv:0705.0663 [hep-th]].
  %%CITATION = JHEPA,0706,054;%%

%\cite{Beccaria:2007pb}
\bibitem{Beccaria:2007pb}
  M.~Beccaria,
  {\em Three loop anomalous dimensions of twist-3 gauge operators in ${\cal N}=4$ SYM}, 
  JHEP {\bf 0709}, 023 (2007)
  [arXiv:0707.1574 [hep-th]].
  %%CITATION = JHEPA,0709,023;%%


%\cite{Beccaria:2008fi}
\bibitem{Beccaria:2008fi}
  M.~Beccaria and V.~Forini,
  {\em Reciprocity of gauge operators in ${\cal N}=4$ SYM}
  arXiv:0803.3768 [hep-th].
  %%CITATION = ARXIV:0803.3768;%%

%\cite{Frolov:2002av}
\bibitem{Frolov:2002av}
  S.~Frolov and A.~A.~Tseytlin,
  {\em Semiclassical quantization of rotating superstring in $\ads$}, 
  JHEP {\bf 0206}, 007 (2002)
  [arXiv:hep-th/0204226].
  %%CITATION = JHEPA,0206,007;%%

%\cite{Kruczenski:2006pk}
\bibitem{Kruczenski:2006pk}
  M.~Kruczenski, J.~Russo and A.~A.~Tseytlin,
  {\em Spiky strings and giant magnons on $S^5$}, 
  JHEP {\bf 0610}, 002 (2006)
  [arXiv:hep-th/0607044].
  %%CITATION = JHEPA,0610,002;%%

%\cite{Ishizeki:2007we}
\bibitem{Ishizeki:2007we}
  R.~Ishizeki and M.~Kruczenski,
  {\em Single spike solutions for strings on $S^2$ and $S^3$}, 
  Phys.\ Rev.\  D {\bf 76}, 126006 (2007)
  [arXiv:0705.2429 [hep-th]].
  %%CITATION = PHRVA,D76,126006;%%

%\cite{Ishizeki:2007kh}
\bibitem{Ishizeki:2007kh}
  R.~Ishizeki, M.~Kruczenski, M.~Spradlin and A.~Volovich,
  {\em Scattering of single spikes}, 
  JHEP {\bf 0802}, 009 (2008)
  [arXiv:0710.2300 [hep-th]].
  %%CITATION = JHEPA,0802,009;%%

%\cite{Kruczenski:2008bs}
\bibitem{Kruczenski:2008bs}
  M.~Kruczenski and A.~A.~Tseytlin,
  {\em Spiky strings, light-like Wilson loops and pp-wave anomaly}, 
  arXiv:0802.2039 [hep-th].
  %%CITATION = ARXIV:0802.2039;%%

%\cite{Derkachov:1999pz}
\bibitem{Derkachov:1999pz}
  S.~E.~Derkachov,
  {\em Baxter's $Q$-operator for the homogeneous $XXX$ spin chain}, 
  J.\ Phys.\ A  {\bf 32}, 5299 (1999)
  [arXiv:solv-int/9902015].
  %%CITATION = JPAGB,A32,5299;%%

\bibitem{Bax72}
  R.J. Baxter,
  Annals Phys. 70 (1972) 193;
  {\em Exactly Solved Models in Statistical Mechanics},
  Academic Press (London, 1982).
  %%CITATION = APNYA,70,193;%%

\bibitem{KorTrick1}
%\cite{Korchemsky:1995be}
%\bibitem{Korchemsky:1995be}
  G.~P.~Korchemsky,
  {\em Quasiclassical QCD pomeron},
  Nucl.\ Phys.\  B {\bf 462}, 333 (1996)
  [arXiv:hep-th/9508025].
  %%CITATION = NUPHA,B462,333;%%


%\cite{Derkachov:2002tf}
\bibitem{Derkachov:2002tf}
  S.~E.~Derkachov, G.~P.~Korchemsky and A.~N.~Manashov,
  {\em Separation of variables for the quantum $SL(2,\mathbb{R})$ spin chain},
  JHEP {\bf 0307}, 047 (2003)
  [arXiv:hep-th/0210216].
  %%CITATION = JHEPA,0307,047;%%

%\cite{Derkachov:1999ze}
\bibitem{Derkachov:1999ze}
  S.~E.~Derkachov, G.~P.~Korchemsky and A.~N.~Manashov,
  {\em Evolution equations for quark gluon distributions in multi-color QCD  and open spin chains}
  Nucl.\ Phys.\  B {\bf 566}, 203 (2000)
  [arXiv:hep-ph/9909539].
  %%CITATION = NUPHA,B566,203;%%

%\cite{Beisert:2003tq}
\bibitem{Beisert:2003tq}
  N.~Beisert, C.~Kristjansen and M.~Staudacher,
  {\em The dilatation operator of ${\cal N} = 4$ super Yang-Mills theory}, 
  Nucl.\ Phys.\  B {\bf 664}, 131 (2003)
  [arXiv:hep-th/0303060].
  %%CITATION = NUPHA,B664,131;%%

%\cite{Korchemsky:1992xv}
\bibitem{Korchemsky:1992xv}
  G.~P.~Korchemsky and G.~Marchesini,
  {\em Structure function for large $x$ and renormalization of Wilson loop}, 
  Nucl.\ Phys.\  B {\bf 406}, 225 (1993)
  [arXiv:hep-ph/9210281].
  %%CITATION = NUPHA,B406,225;%%

%\cite{Belitsky:2003ys}
\bibitem{Belitsky:2003ys}
  A.~V.~Belitsky, A.~S.~Gorsky and G.~P.~Korchemsky,
  {\em Gauge / string duality for QCD conformal operators}, 
  Nucl.\ Phys.\  B {\bf 667}, 3 (2003)
  [arXiv:hep-th/0304028].
  %%CITATION = NUPHA,B667,3;%%

%\cite{Belitsky:2006en}
\bibitem{Belitsky:2006en}
  A.~V.~Belitsky, A.~S.~Gorsky and G.~P.~Korchemsky,
  {\em Logarithmic scaling in gauge / string correspondence}, 
  Nucl.\ Phys.\  B {\bf 748}, 24 (2006)
  [arXiv:hep-th/0601112].
  %%CITATION = NUPHA,B748,24;%%

%\cite{Eden:2006rx}
\bibitem{Eden:2006rx}
  B.~Eden and M.~Staudacher,
  {\em Integrability and transcendentality}, 
  J.\ Stat.\ Mech.\  {\bf 0611}, P014 (2006)
  [arXiv:hep-th/0603157].
  %%CITATION = JSTAT,0611,P014;%%

%\cite{Beisert:2006ez}
\bibitem{Beisert:2006ez}
  N.~Beisert, B.~Eden and M.~Staudacher,
  {\em Transcendentality and crossing}, 
  J.\ Stat.\ Mech.\  {\bf 0701}, P021 (2007)
  [arXiv:hep-th/0610251].
  %%CITATION = JSTAT,0701,P021;%%

%\cite{Freyhult:2007pz}
\bibitem{Freyhult:2007pz}
  L.~Freyhult, A.~Rej and M.~Staudacher,
  {\em A Generalized Scaling Function for AdS/CFT},
  arXiv:0712.2743 [hep-th].
  %%CITATION = ARXIV:0712.2743;%%

\bibitem{Grob}
T.~Becker, V.~Weispfenning, with H.~Kredel,
{\em Gröbner Bases: A Computational Approach to Commutative Algebra},
Springer; 1st ed. 1993. Corr. 2nd printing edition (March 23, 1998).

\end{thebibliography}
\end{document}